\documentclass[twocolumn,showpacs,prd]{revtex4}
\usepackage{amsfonts, amssymb,amsmath} 
\usepackage{graphicx}
\usepackage{dcolumn}
\usepackage{bm}

\newcommand{\be}{\begin{equation}}
\newcommand{\ee}{\end{equation}}
\newcommand{\bea}{\begin{eqnarray}}
\newcommand{\eea}{\end{eqnarray}}

\newcommand{\tc}{\c c}

\newcommand{\tg}{\u g}

\newcommand{\tI}{\.I}

\begin{document}

\title{EXPONENTIALLY EXPANDING RADIATION DOMINATED AND DUST DOMINATED UNIVERSES IN BRANS-DICKE THEORY}
\author{M. Ar\i k$^1$ and T. Erkmen$^1$}
\affiliation{$^{1}$Department of Physics, Bo\tg azi\tc
 i University, Bebek, 34342, \tI stanbul, Turkey}
\date{\today}

\begin{abstract}
The Brans-Dicke Theory of Gravity is one of the most promising
alternatives
 to the Einstein's Theory of General Relativity.
  We have examined an action term with wrong signs for both the  kinetic and mass
  terms for the scalar field
 and have found solutions for both the scale factor of the universe and the Brans-Dicke scalar
field which vary exponentially in time.

\end{abstract}
\pacs{98.80.Jk, 04.20.Jb, 04.50.+h, 95.36.+x}
\maketitle

Recent observations of Type Ia supernovae [1]-[2] and measurements
of the cosmic microwave background [3] suggests that the universe
is in an accelerating expansion phase [4]. The universe seems to
be flat [5]-[7], with energy content, $\Omega_{dust}=0.25$ which
contains baryonic matter and CDM with pressure $P_{dust}=0$
[8]-[11]. The universe seems predominantly filled with dark energy
or quintessence which drives the expansion by its negative
pressure [12]-[14]. In this paper we will assume that the universe
is filled homogenously by quintessence, the scalar field, $\phi$.

The Brans-Dicke Theory of Gravity is one of the leading
alternatives to Einstein's Theory of General Relativity [15]. It
may be said that Einstein's Theory of General Relativity (GR) is a
special case of Brans-Dicke Theory. As the dimensionless constant
$\omega$, of Brans-Dicke Theory approaches infinity, the theory
approximates GR [16].

We will model the universe with Brans-Dicke Theory with the scale
factor varying exponentially with time, that is $a(t)=e^{Ht}$. Let
us consider a simple minded model where the relationship between
the scalar field $\phi$ and the energy density of rest of the
matter $\rho_M$ is given as,
\begin{eqnarray}
\frac{1}{2}\dot{\phi}^2+\frac{1}{2}m^2{\phi}^2=\rho_M
 .
\end{eqnarray}
We are assuming the quintessence, the scalar field tracks the
matter energy density $\rho_M$ according to this equation. If
there's an exponential solution for the scalar field such that
$\phi(t)=e^{Ft}$ the above equation will become
\begin{eqnarray} \label{phirho}
\frac{1}{2}\left(F^2+m^2\right)\phi^2=\rho_M .
\end{eqnarray}
We will have a linear relationship between the $\phi^2$ and
$\rho_M$ terms. We further assume the scale factor of the universe
varies exponentially with respect to time, $a(t)=e^{Ht}$. For dust
dominated universe $\rho_{dust}\sim{1}/{a^3}$, so we get,
\begin{eqnarray}
\rho_{dust}\sim\frac{1}{a^3}=e^{-3Ht} .
\end{eqnarray}
From ($\ref{phirho}$) we get
\begin{eqnarray}
\phi\sim e^{-{3}/{2}Ht} ,
\end{eqnarray}%
so F is equal to,
\begin{eqnarray}
F=-{3}/{2}H .
\end{eqnarray}
Similarly for the radiation dominated universe
$\rho_{radiation}\sim{1}/{a^4}$, so we get,
\begin{eqnarray}
\rho_{radiation}\sim\frac{1}{a^4}=e^{-4Ht}
\end{eqnarray}
and
\begin{eqnarray}
\phi\sim e^{-2Ht} ,
\end{eqnarray}%
\begin{eqnarray}
F=-2H .
\end{eqnarray}
We will see that our rigorous results obtained using Brans-Dicke
theory will match these values. We thus consider the action
\begin{eqnarray}
S = \int d^{4}x\sqrt{g}\left(-\frac{1}{8\omega}\phi^{2}R
-\frac{1}{2}g^{\mu\nu}{\partial}_\mu\phi{\partial}_\nu\phi+\frac{1}{2}m^{2}\phi^{2}+{L}_M\right)
  .
\end{eqnarray}
The signature of the metric is (+,-,-,-). The kinetic and the mass
terms of the scaler field are negative of the action of the
standard Brans-Dicke Theory, i.e. in field theory terminology the
$\phi$-field is a ghost. However since both the signs of the
kinetic and the mass terms are changed the free $\phi$ field
satisfies the conventional Klein-Gordon equation with plane wave
solutions. The fact that there are ghost fields in a theory does
not mean that the theory is unphysical. As an example one may
mention the Lee-Wick model which includes ghost fields but has
unitary $S$ matrix [17].

The variation of this action with respect to Robertson Walker
metric gives us the following gravitational field equations;
\begin{equation}
\frac{3}{4\omega}\,\phi^{2}\,\left(  \frac{\dot{a}^{2}}{a^{2}}+\frac{k}{a^{2}%
}\right)
+\frac{1}{2}\,\dot{\phi}^{2}+\frac{1}{2}\,m^{2}\,\phi^{2}+\frac
{3}{2\omega}\,\frac{\dot{a}}{a}\,\dot{\phi}\,\phi=\rho_M\label{des}%
\end{equation}%
\begin{equation}
\frac{-1}{4\omega}\phi^{2}\left(  2\frac{\ddot{a}}{a}+\frac{\dot{a}^{2}}%
{a^{2}}+\frac{k}{a^{2}}\right)  -\frac{1}{\omega}\,\frac{\dot{a}}{a}%
\,\dot{\phi}\,\phi-\frac{1}{2\omega}\,\ddot{\phi}\,\phi+
\nonumber
\end{equation}%
\begin{equation}
+\left(  \frac{1}%
{2}-\frac{1}{2\omega}\right)
\,\dot{\phi}^{2}-\frac{1}{2}\,m^{2}\,\phi^{2}=P_M
\label{pres}%
\end{equation}%
\begin{equation}
\ddot{\phi}+3\,\frac{\dot{a}}{a}\,\dot{\phi}+\left[
m^{2}+\frac{3}{2\omega }\left(
\frac{\ddot{a}}{a}+\frac{\dot{a}^{2}}{a^{2}}+\frac{k}{a^{2}}\right)
\right]  \,\phi=0 \label{fi}%
  ,
\end{equation}
where k is the curvature parameter with k= -1, 0 and 1
corresponding to open, flat and closed universes respectively.
$a(t)$ is the scale factor of the universe. Dot denotes derivative
with respect to time. M stands for everything except the
Brans-Dicke scalar field.

We will solve the above equations for flat universe $k=0$, and
with the scale factor of the universe $a$ and the scalar field
$\phi$ varying exponentially with time. That is; $a(t)=e^{Ht}$ and
$\phi(t)=e^{Ft}$ . The scalar field, $\phi$ is independent of
space coordinates, it is homogeneously distributed over the
universe. The three gravitational field equations above
(\ref{des}, \ref{pres}, and \ref{fi}) will simply give;

\begin{equation} \label{phiHFm}
-F^{2}-3HF-\frac{3}{\omega}H^{2}=m^{2}\\
\end{equation}
\begin{equation}  \label{phiHFmrho}
\frac{3}{4\omega}H^{2}+\frac{F^2}{2}+\frac{m^2}{2}+\frac{3}{2\omega}HF=\frac{{\rho}_M}{\phi^{2}}
\end{equation}
\begin{equation}   \label{phiHFmP}
-\frac{3}{4\omega}H^{2}-\frac{HF}{\omega}-\frac{F^2}{\omega}+\frac{F^2}{2}-\frac{m^2}{2}=\frac{P_M}{\phi^{2}}
  .
\end{equation}

By substituting (\ref{phiHFm}) into (\ref{phiHFmrho}), and
(\ref{phiHFmP}) we get,

\begin{equation}  \label{phiHFrho}
-\frac{3}{4\omega}H^{2}-\frac{3}{2}HF\left(\frac{\omega-1}{\omega}\right)=\frac{{\rho}_M}{\phi^{2}}
\end{equation}
\begin{equation}   \label{phiHFP}
\frac{3}{4\omega}H^{2}+\frac{3}{2}HF\left(\frac{\omega-1}{\omega}\right)+\frac{1}{2\omega}HF+F^{2}\left(\frac{\omega-1}{\omega}\right)=\frac{P_M}{\phi^{2}}
  .
\end{equation}

For ${\rho}_M > 0 $  (\ref{phiHFrho}) gives

\begin{equation} \label{const1}
-F>\frac{H}{2(\omega-1)}
\end{equation}
this constraint (\ref{const1}) also satisfies $P_M+{\rho}_M > 0 $
when substituted into (\ref{phiHFrho}) and (\ref{phiHFP}).

$\nu={P_M}/{{\rho}_M}$ can be obtained as;

\begin{equation}
\nu=\frac{P_M}{{\rho}_M} = -1+\frac{2}{3}\frac{(-F)}{H}
  .
\end{equation}

Notice that $\nu$ is independent of the Brans-Dicke parameter
$\omega$. Since we have assumed an exponential expansion for the
universe, in the Einsteinian limit where the gravitational
constant does not change, $F=0$ and one obtains $\nu=-1$ as usual.
However for the general Brans-Dicke case one can have exponential
expansion for any value of $\nu$ provided that the gravitational
constant also changes according to this equation.

The ratios of ${H}/{m}$ and ${F}/{m}$ can be found as
\begin{equation}
\frac{2} {3\sqrt{-\nu^2+\left(1-\frac{4}{3\omega}\right) }} =
\frac{H}{m}
\end{equation}
and
\begin{equation}
\frac{\nu+1} {\sqrt{-\nu^2+\left(1-\frac{4}{3\omega}\right) }}
 = \frac{-F}{m}
\end{equation}
for $\nu$ sufficiently larger than -1 and sufficiently smaller
than 1. That is, $-\sqrt{1-4/3{\omega}}<\nu<\sqrt{1-4/3{\omega}}$.
Time delay experiments give the present lower limit for the
constant $\omega$  as $\omega>10^4$ [18]-[20]. So
${1}/{\omega}\ll1$, the ${1}/{\omega}$ terms may be ignored.

For a radiation dominated universe $\nu={1}/{3}$ so that $
H={m}/{\sqrt{2}} $ and $ F=-\sqrt{2}m$. Whereas for a dust
dominated universe $\nu=0$ so that $ H={2}/{3}m $ and $ F=-m$.

We see that the Hubble constant slightly decreases as the universe
evolves from the radiation dominated era into the matter dominated
era. In fact the minimum value of H is obtained for $\nu=0$. The
Newtonian gravitational constant $ G_N \sim \phi^{-2} \sim
e^{-2Ft} $, on the other hand increases at a faster rate during
radiation dominated era as $-F$ increases as $\nu$ increases. Note
that in this type of model ${\dot{G}_N}/{G_N}$ is positive and is
of order of Hubble constant. This is in contrast to more
conventional models where ${\dot{G}_N}/{G_N}$ is negative and is
order of ${1}/{\omega}$ times the Hubble constant [21]. Thus this
type of model predicts a time varying Newton's gravitational
constant which is more amenable to being tested by experiment.

\begin{figure}
\centering
\includegraphics[angle=0,width=0.5\textwidth,height=0.40\textwidth]{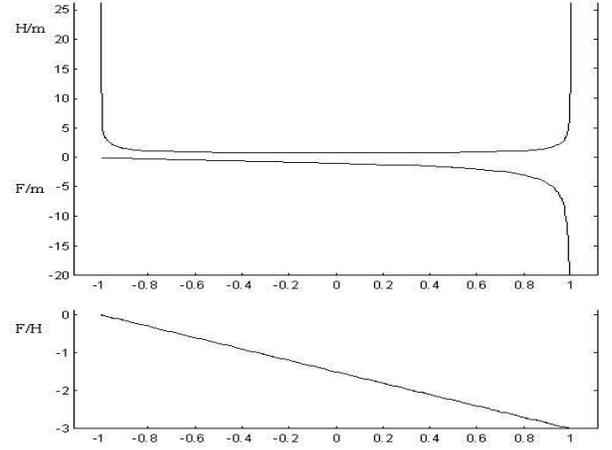}
 \caption{H/m, F/m and F/H versus $\nu$.}
\end{figure}

This work is supported by Bo\u gazi\c ci University Research Fund,
Project no: 07B301.

\end{document}